# XANES analysis of structural changes in a transient photoexcited state of metalloporphyrin

S. Della Longa[1], L. X. Chen[2], P. Frank[3], K. Hayakawa[4,5], K. Hatada[4], and M. Benfatto[4]

[1] Dip. Medicina Sperimentale, Università dell' Aquila, 67100 L'Aquila, Italy.

[2] Chemical Sciences and Engineering Division, Argonne National Laboratory, 9700 S. Cass Ave., Argonne, Illinois, 60439 USA

[3] Department of Chemistry, Stanford University, Stanford, CA 94305-5080, USA

[4] Laboratori Nazionali di Frascati, LNF- INFN, Frascati, Italy

[5] Museo Storico della Fisica e Centro Studi e Ricerche ``Enrico Fermi'', Via Panisperna 89A, 00184 Roma, Italy

# ABSTRACT

We have performed a structural analysis of the Ni K-edge XANES spectrum of a square planar coordination complex, Ni-tetramesitylporphyrin, Ni(II)TMP, in dilute toluene solution. The fit of the spectrum was carried out in the frame of the full multiple scattering (FMS) approach, via the MXAN program (Benfatto & Della Longa, 2001), starting from a muffin-tin (MT) form of the molecular potential.

We show that, by excluding pre-edge and lower rising edge features from the fit, an excellent simulation and structural recovery can be obtained for the compound under study. According to our XANES analysis, the first-shell and second-shell distances are, respectively, d(Ni-N)=(1.93±0.02) Å and d(Ni-C)=(2.94±0.03) Å, in agreement with previous EXAFS results (L.X. Chen et al. JACS 2007, 129:9616-9618).

Furthermore, we have applied the MXAN analysis to the time-resolved XANES difference spectrum obtained between a 100 ps life-time photoexcited state of Ni(II)TMP (a relaxed triplet state, $T_1$, obtained by exciting the Soret band) and the ground state, $S_0$ (Chen et al. 2007). From our direct simulation of the laser on/laser off difference spectrum, we obtain the fraction of $T_1$ population (0.41-0.63), in agreement with optical transient absorption determination, and a 0.05-0.1 Å elongation of the d(Ni-N) and d(Ni-C) distances. The overall uncertainty on these results mostly depends on uncertainty on the chemical shift between the $T_1$ and $S_0$ state, that is assumed "a priori" in the differential analysis. Still our results are in good agreement with previous EXAFS data (obtained on a reconstructed absolute spectrum of the $T_1$ state). Our best fit corresponds to no chemical shift, 54% fraction of $T_1$ state, and either a (0.04±0.01) elongation of the average d(Ni-N) or a (0.06±0.01) elongation of the average d(Ni-pyrrol) when pyrrol rings are treated as perfectly rigid bodies.

This work demonstrates that XANES spectroscopy can be used to provide structural information on square-planar transition metal compounds by analysing the metal K-edge via MS method under the MT approximation, and that XANES differential analysis can be successfully applied to study electronic/structural relationships in transient photoexcited species from time-resolved experiments.

# INTRODUCTION

X-ray Absorption spectroscopy (XAS) is a powerful tool for determining both electronic and structural information on the system of interest. On one hand, X-ray Absorption Near-Edge Structure (XANES) around the transition edge provides details about the electronic structure of the atom of interest, i.e., orbital occupancy, oxidation state, and ligand field strength and coordination symmetry of its valence states. On the other hand, coordination numbers and bond distances around a specific atom can be determined from the Extended X-ray Absorption Fine Structure (EXAFS). Hence, XAS can be particularly useful in search of relationships between electronic states and structural geometry for metal centers in chemistry, biology as well as materials sciences.

In recent years, there has been growing the number of studies about transient photo-excited states by XAS spectroscopy using a pump-probe scheme where the system is excited by an ultrashort pump laser pulse and is probed by a time-delayed X-ray pulse (Chen et al., 2001; Chen et al; 2002a; Chen, 2002b; Chen et al., 2003; Saes et al. 2003a, 2003b, 2004; Bressler et al. 2004; Chen, 2004; Jennings et al., 2002; Chen et al., 2004; Chen, 2005; Gawelda et al. 2005, Majed et al 2006; Chen et al, 2007). In this way it is possible to obtain both electronic and structural changes due to the photoexcitation. Metalloporphyrins are versatile substrates in various photochemical reactions, enabling intra- and intermolecular energy or electron redistribution that triggers chemical processes. However, details on electronic configuration and molecular geometry of photoexcited state are not completely understood from previous steady-state structural determination methods or indirect structural methods, such as optical transient absorption spectroscopy.

Several years ago Benfatto and Della Longa (Benfatto and Della Longa, 2001) proposed a fitting procedure, MXAN (Minuit XANES) based on a full multiple scattering (MS) theory, which could extract local structural information around the absorbing atom from experimental XANES data. Since then, the MXAN method has been successfully used for analyses of many known and unknown systems, and the numerical results have been compared with X-ray Diffraction and/or EXAFS results (Benfatto et al. 2002; D' Angelo et al. 2002; Hayakawa et al. 2004; Frank et al. 2005; Arcovito et al. 2006).

Recently the MXAN method has been modified to enable the extraction of structural information from a difference spectrum obtained by subtracting the ground state spectrum from the spectrum obtained under an external perturbation, such as the photoexcitation by a laser pulse. The importance of differential XAS spectra for revealing very small structural and electronic changes in a sample has been emphasized in both EXAFS and XANES regions recent years (Pettifer et al. 2005; Arcovito et al 2005; Benfatto et al. 2006). This method has advantages of enhanced sensitivity of curve fitting and reduced influence of possible systematic errors in calculations due to the approximations used in the theoretical approach. Key points underlying the validity of this kind of analysis are a perfect energy stability during the experiment, and a proper knowledge of some possible chemical edge-shift (Pettifer et al 2005).

**In this paper we present an analysis in the XANES region of Ni K-edge for a metalloporphyrin, nickeltetramesitylporphyrin (NiTMP), in a dilute toluene solution. Our analysis is based on the experimental XANES spectra of NiTMP in its ground and photoexcited states, respectively. The latter was obtained within 100 ps of the laser pulse excitation of the Q-band at 527 nm. The experimental data, shown in Fig. 1, were already reported by Chen et al.** (2007) where an EXAFS analysis of the $S_0$ spectrum and of the $T_1$ spectrum was performed. The $T_1$ state spectrum was reconstructed by subtracting the remaining $S_0$ fraction (measured by the optical transient absorption spectroscopy) from the spectrum with the laser excitation.

Our aim is to extract the structure of the $S_0$ and $T_1$ states by XANES spectra, without using an additional information on the $T_1$ state fraction, because this quantity can be obtained with our direct analysis of the difference spectrum [With the laser excitation – No laser excitation (the $S_0$ state)]. Meanwhile, we want to use this example to validate the MXAN method for applications to K-edge XANES spectra of low coordination symmetry transition metal complexes, which normally have richly featured rising edge region. However, several features in the first 10 eV above the transition edge are difficult to model in detail due to 1) the breakdown of the muffin-tin (MT) approximation at very low energy in the open system, 2) the presence of shake-up (shake-off) processes, and/or 3) contributions of distant neighboring atoms that are not directly coordinated to the absorbing atom. In this report, we will show how MXAN method can successfully treat this type of systems with an excellent simulation and local structural recovery that agree with previous EXAFS studies. Meanwhile, our results also demonstrate that the differential XANES spectral analysis can be successfully applied to analyzing time-resolved transient XAS data in getting a complete local structural characterization for the transient state.

**METHODS**

All details on the x-ray transient absorption (XTA) experiment with ~100 ps time resolution were reported previously (Chen et al. 2007). Emphasized here is the data acquisition method that allowed intrinsic energy calibration between spectra of the ground state and the laser excited mixture state. The energy calibration of the experiment was first carried out extrinsically with the Ni foil standard. Then the energy calibration was also performed intrinsically, by taking the data for the ground and the laser excited states alternately at each data point. This intrinsic energy calibration was accomplished by gating a nine-element solid state germanium detector array (Canberra) after the signals were transformed through the amplifiers and single channel analyzers (SCA). The signals were then connected to two sets of scalers with two gate signals with the same repetition rate as the laser pump pulse at 1kHz, but at different time delays from the laser excitation, one of which was coincide with the laser pulse and used for the spectrum of photoexcited sample, and the other of which was at 147.2 μs later after the laser pulse and used for the ground state spectrum. Hence, any edge energy change from the ground state was resulted from the structural change due to the laser excitation, rather than experimental error in energy calibration.

**Coordination Models**. We analysed the effects of porphyrin macrocycle distortions on the XANES spectra starting from two idealized porphyrin macrocycle structures, the former having a perfect square-planar conformation, the latter having a saddled macrocycle with the metal atom as the saddle point. The coordinates for the former were taken from those of 5,10,14,20-tetramesitylporphyrinato-copper(ii) hexachloroantimonate dichloromethane solvate (Reed & Scheidt, 1989), replacing Cu(II) with Ni(II). The coordinates of the saddled conformation were taken from those of 2,3,7,8,12,13,17,18-Octabromo-5,10,15,20-tetramesityl-porphyrinato)-Ni(II) (Mandon et al., 1992).

**XAS theory and minimization procedure.** Calculation and fitting of XANES spectra has been carried out via the MXAN program (Benfatto & Della Longa 2001) that is based on the full multiple scattering (MS) approach in the framework of the muffin tin (MT) potential approximation. It also takes into account inelastic processes by way of a Lorentzian broadening function. This approach has been successfully applied to K-edge spectra of a number of transition metal compounds (Benfatto et al. 2002; D' Angelo et al. 2002; Hayakawa et al. 2004; D' Angelo et al. 2006), and also to interpret XANES spectra taken on hemeproteins in crystal state (Della Longa et al. 2001; Della Longa et al. 2003; Arcovito et al. 2007), and solution (Arcovito et al. 2005).

Details on the MXAN method when applied to differential XANES spectra [photoexcited state – ground state] were give previously, at the Fe K-edge of the cryotrapped photolysis product of carbonmonoxy-myoglobin, upon illumination at 15K (Arcovito et al 2005), and at the Ru L-edge and Fe K-edge upon photoexcitation of aqueous $[Ru^{II}(bpy)_3]^{2+}$ and $[Fe^{II}(bpy)_3]^{2+}$, respectively, in picosecond time-resolved experiments (Benfatto et al. 2006, Gavelda et al. 2007). Minimization of the $\chi^2$ function was performed in the $[R_1,R_2]$ parameter space, where $R_1= d(Ni-N_p)$ and $R_2= d(Ni-C)$ are the first shell and second-shell distances, respectively. Alternatively, the pyrrol rings of the porphyrin was treated as a perfectly rigid body, and fitting procedures were performed in the single parameter space of variable $R_1$.

Energy scale in fitting figures is the theoretical one, the zero-energy (the continuum level $V_0$) corresponding to $(8341.9\pm0.3)$ eV

**Evaluation of the $T_1$ fraction from the MXAN procedure**: The MXAN program calculates the absolute cross section $\sigma_{abs}(E)$ of the X-ray absorption edge of the metal cluster under study, the physical units being MBarn. A proportional constant $A_{norm}$ is used to scale the calculated $\sigma_{abs}(E)$ to the measured data $\mu_{abs}(E)$. Of course the numerical values of $\mu_{abs}(E)$ and in turn the value of $A_{norm}$ can vary depending on metal concentration and physical units used. Usually spectra of $\mu_{abs}(E)$ are given in units where the edge jump is equal to 1.

In order to evaluate the $T_1$ fraction from our XANES difference spectrum, two assumptions have been made. The first one is that the constant $A_{norm}$ is the same for the $S_0$ and $T_1$ experimental spectra:

$$\sigma_{S_0}(E) = A_{norm}\mu_{S_0}(E)$$

$$\sigma_{T_1}(E) = A_{norm}\mu_{T_1}(E)$$

Even if the $T_1$ spectrum is not directly measured, this is certainly true in our case because $S_0$ and $T_1$ states refer to the same sample in the same experimental conditions. The second assumption is that the difference spectrum refers to a perfect two-state transition. In this case if only a fraction $x$ of $S_0$ converts to the $T_1$ state, the overall shape of the XANES difference spectrum will be identical to the spectrum corresponding to 100% $T_1$ conversion, and the amplitude ratio between them will be given by the $T_1$ fraction $x$. In other words, under these quite reasonable assumptions, for the difference spectrum $[T_1 - S_0]$ corresponding to 100% conversion:

$$\sigma_{T_1}(E) - \sigma_{S_0}(E) = A_{norm}[\mu_{T_1}(E) - \mu_{S_0}(E)]$$

**whereas for the one corresponding to conversion of the $T_1$ fraction $x$:**

$$\sigma_{T_1}(E) - \sigma_{S_0}(E) = A'_{norm}[\mu_{T_1}(E) - \mu_{S_0}(E)]$$

**where $A'_{norm} = xA_{norm}$. Thus, the $A/A'$ ratio gives the $T_1$ fraction $x$, $A$ and $A'$ being extracted from the MXAN procedure when applied to the $S_0$ absolute spectrum, and to the difference spectrum, respectively.**

## RESULTS

**Fig. 1 displays the experimental Ni K-edge XANES spectra of Ni(II)TMP at its ground state, $S_0$, and the laser excited mixture taken at 100 ps delay after the laser pump pulse (1 mJ, 5 ps FMHM, 527 nm, and 1 kHz repetition rate), as reported by Chen et al. 2007. According to the result from the optical transient absorption measurements, the mixture contained 45-50% of the excited state, $T_1$. Weak features 1 (at 8331 eV) and 2 (at 8333.2) are present in the pre-edge region, and are assigned to $1s \rightarrow 3d$ transitions that are sensitive to electronic structural change induced by the laser.** The rising edge of the spectra exhibits a distinctive sharp feature **3** (at 8338 eV), which is shifted to 1.5 eV higher in energy in the $T_1$ state than in the $S_0$ state. The origin of such a shift is not completely clear, although the result agrees with the electron density reduction of Ni(II) due to the ring expansion observed in the EXAFS spectra.(Chen, et al. 2007) Moreover, significant changes extended to the "white line" peak region of 8350-8360 eV (peaks **4** and **5**) and beyond (peaks **6-10**) are also observed between the two spectra, which have been attributed to energy level shifts of $4p_x$ and $4p_y$ MOs convoluted with multiple scattering contributions. The energy splitting between **peaks 3 and 4, and between peaks 4 and 5, are about 14 eV and 5 eV, respectively. In the lower panel of Fig. 1, the XANES difference spectrum of the same states, [photoexcited – $S_0$] is displayed. By assuming that the observed spectral changes are due to a perfect two-state transition where only a fraction of $S_0$ converts to the $T_1$ state, the overall shape of the XANES difference spectrum will be identical to the spectrum corresponding to 100% $T_1$ conversion, the overall amplitude being given by the $T_1$ fraction.**

**XANES calculations of the $S_0$ state.** Given these experimental results, we started calculating the XANES spectrum of a ground state Ni(II)porphyrin cluster as a function of the cluster size. The calculated XANES spectra corresponding to 1-shell (NiN$_4$), 2-shell (NiN$_4$C$_{12}$) and 3-shell (Ni-porphyrin) clusters are depicted in Fig.2, respectively. It is evident that all the experimental features **4** to **10** are well reproduced by our calculations. We note that including three shells is essential for a successful simulation. In the inset, a blow-up of the edge features in the 3-shell calculation is displayed. The calculated edge contains a weak shoulder (feature **3'**) that does not correspond to the experimental peak **3**, having a different energy splitting from peak **4** (8.5 instead of 14 eV). Possible explanations of the failure in reproducing peak **3** are given in the discussion section.

We applied the MXAN procedure to fit the XANES spectrum of the ground state, $S_0$ (Fig. 3). Although we cannot reproduce peaks **1**, **2** and **3,** the fit is satisfactory in the energy range from above peak **3** to 8550 eV (k = 7 A$^{-1}$). Hence, our ability of measuring structural parameters of a metalloporphyrin is verified though the energy range for the fitting is slightly shorter (by about 5-10 eV) than that in previous studies on metal ions in water and hemeproteins (see literature in the Methods). We fit the spectrum in the [R$_1$,R$_2$] parameter space, where R$_1$= d(Ni-N$_p$) and R$_2$= d(Ni-C) are the first shell and second-shell distances, respectively.

**We obtained R$_1$=(1.93±0.02) and R$_2$=(2.94±0.03), that is consistent with EXAFS values R$_1$=1.90 and R$_2$=2.92 (Chen et al, 2001, 2007). We note here that R$_1$ and R$_2$ values remain unchanged by including peaks 1, 2 and 3** **in the fit, but the reduced $\chi^2$ from the fit increases by about 400% due to the failure in reproducing peak 3. However these results are clear evidence that the MT approach can be successfully applied to open systems like the square-planar Ni porphyrin, giving an excellent theoretical reproduction and quantitative structural recovery.**

**XANES analysis of the (photoexcited – $S_0$) difference spectrum**. In order to get further insight into the structure of the excited state $T_1$, we have applied MXAN to directly fit the experimental XANES difference spectrum [photoexcited – $S_0$] to obtain the optimized average $d(Ni-N_p)$ and $d(Ni-C)$. This analysis is different from the EXAFS analysis performed by Chen et al. 2007, that was carried out on a reconstructed $T_1$ spectrum obtained by a linear combination of the $S_0$ and [photoexcited – $S_0$] experimental spectra, imposing the $T_1$ fraction evaluated from the transient optical spectroscopy data. In our analysis the normalization constant between the theoretical and experimental difference spectrum is allowed to vary. As explained in the Methods, under the assumption of a perfect two-state transition, the ratio between the optimised normalization constant of the difference spectrum [photoproduct – $S_0$] and the normalization constant of the absolute spectrum of $S_0$, allows us to directly estimate the $T_1$ fraction.

When considering XANES difference spectra, a chemical shift between the two spectra can occur, reflecting redox processes that alter the net charge of the metal center. This effect does not come out from our calculations, and have to be included "a priori" in the analysis. Finally, differential XAS analysis is extremely sensitive to energy calibration of the beam line, that has to be checked very carefully. As stated in the Method section, both extrinsic and intrinsic energy calibrations were implemented in the data acquisition. Therefore, the edge energy differences between $S_0$ and $T_1$ states are due to the electronic structural change not due to the systematic error.

Results from a set of fits are given in Table 1, each of which differs from others by the choice of structural variables, and the imposed chemical shift. **Fits #1 – #5 are performed in the double variable space [$R_1$, $R_2$], while fits #6 – #8 are performed in the single variable space d(Ni-$N_p$), and the pyrrol ring is considered a perfectly rigid ring. The best fits for the XANES spectrum are fits #1 and #6 shown in Fig. 4. The $T_1$ fraction is extracted from the fits as mentioned above. Fits were repeated in the double and single parameter space assuming difference chemical shift values ranging from 0.0 to -2.0 eV, with the negative sign corresponding to a blue-shift of the XANES spectrum relative to that of $S_0$, indicating more positive charges on Ni in the $T_1$ state. This set of choices in chemical shifts matches the experimental observation of the $T_1$** state spectrum with blue-shifts in peak 2 by 0.4 eV, and peak 3 by 1.5 eV, respectively (Chen et al. 2007).

The best fit (#1) corresponds to a chemical shift of 0.0 eV, the $T_1$ state fraction of 54%, $\Delta(R_1)=0.04\pm0.02$ Å and $\Delta(R_2)=0.05\pm0.03$ Å, with a $\chi^2/n$ value of 2.21. Assuming the porphyrin macrocycle is perfectly rigid, we obtained $\Delta(R_1)=0.06\pm0.02$ Å and $T_1$ fraction = 59%, with $\chi^2/n$ = 2.56. Overall fitting results are in a good agreement with the EXAFS analysis by Chen et al. (2007) on the reconstructed $T_1$ spectrum where $\Delta(R_1)=0.08\pm0.02$ Å and $\Delta(R_2)=0.07\pm0.02$ Å. As expected, the fitting results are correlated to the chemical shift chosen between $T_1$ and $S_0$. The change in the first shell distance $\Delta(R_1)$ is evaluated between 0.04 and 0.13 Å, and $T_1$ fraction, between 41% to 59%. By assuming a blue shift of 0.5-1.0 eV, our results become identical to the experimental EXAFS analysis, however, the $\chi^2/n$ increases by about 30%.

**Twisting effects.** We have tested effects of porphyrin macrocycle distortions on the XANES spectra by comparing results from two idealized porphyrin macrocycle structures with a perfect square-planar conformation and a saddled macrocycle with Ni atom as the **saddle point. The two structures are depicted in Fig. 5A and 5B.** The coordinates for the former were taken from a copper(II)-tetramesitylporphyrin crystalline complex (Reed & Scheidt, 1989; see Methods) replacing Cu(II) with Ni(II), and those for the saddled conformation were taken from a nickel(II)-tetramesitylporphyrin crystalline complex (Mandon et al. 1992). In the latter reported structure d(Ni-N) has the same value as measured by EXAFS on Ni(II)TMP in toluene solution (Chen et al. 2007), and the porphyrin macrocycle is twisted around with the metal center as the saddle point. The first-shell coordination symmetry in the saddled porphyrin macrocycle deviates from square-planar towards flattened tetrahedral, with an $N_p$-Ni-$N_p$ angle of 170°, where the two $N_p$ atoms are from the two pyrrol rings opposite with each other. The pyrrol rings of the porphyrin macrocycle are furtherly distorted with respect to the xy plane, as well as rotated along the Ni-$N_p$ bond. For a meaningful comparison, the two clusters used in MS calculations include the same 24 porphyrin atoms (N and C atoms up to a distance of 4.5 Å from Ni, and the pyrrol rings are treated as rigid bodies), and identical line broadening terms.

In order to separate the effects of macrocycle twisting from those of bond elongation, we first compare the results for the twisted and the square-planar structures, while keeping d(Ni-N) fixed at 1.91 Å (Fig. 6A). We then compare two square-planar structures with different d(Ni-N), from 1.91 Å to 2.01 Å. According to our calculations, the porphyrin twisting determines second-order effects on the XANES energy range, i.e. small changes of the main peaks 4 and 5. Hence the overall XANES changes observed, beyond peak 3, in the $S_0 \rightarrow T_1$ transition, are mainly due to changes of the average Ni-pyrrol distance.

## DISCUSSION

We have shown that, by excluding pre-edge and lower rising edge features from the fit, an excellent simulation and structural recovery can be obtained for a square planar Ni(II)TMP. According to our XANES analysis, the first- and second-shell distances are in agreement with previous EXAFS results (Chen et al. 2001, 2007). We have applied the MXAN analysis also to the time-resolved XANES difference spectrum obtained between a 100 ps life-time photoexcited state of Ni(II)TMP and the ground state $S_0$. We have obtained the $T_1$ fraction (0.41-0.63), in agreement with optical transient absorption determination, and a 0.05-0.1 Å elongation of the d(Ni-N) and d(Ni-C) distances. The overall uncertainty on these results mostly depends on uncertainty on the chemical shift between the $T_1$ and $S_0$ state, that is assumed "a priori" in the differential analysis. Still our results are in good agreement with previous EXAFS data obtained on a reconstructed absolute spectrum of the $T_1$ state.

As mentioned in the Results section, there is still a lack in the theoretical ability to reproduce XANES features at very low energy. In principle, a complete theoretical reproduction of the spectrum, leading to a complete picture of the electronic/structural state cound be accomplished as long as various approximation would be improved (Rehr & Albers, 2000; Natoli et al., 2003). Among them, the Muffin Tin (MT) approximation of the molecular potential has been widely mentioned as the main limitation affecting the analysis of the low energy range of the spectrum, especially concerning metal sites with an open coordination sphere. Indeed, the statement about the breakdown of the muffin-tin (MT) approximation is often a rather general claim lacking a deeper discussion and distinction between different experimental cases. The present work confirms our previous studies (Hayakawa et al. 2004) on hexa-coordinated transition metal clusters, where the MT approach and the non-MT approach (by the Finite Difference Method, FDM, Joly, 2001; Joly, 2003) were compared. The MT scheme seems to be valid for the overall XANES range except in the very low energy part, i.e. empty bound states with p- character below the edge. The XAS spectra ferro and ferri hexacyanide were measured both on crystals and in aqueous solution. EXAFS and XANES quantitative minimizations agreed with the crystallographic structures reported on these compounds (Figgis et al. J. 1978; Morioka et al. 1985). We also reported the FDM calculation on these compounds, demonstrating that the improvement in going from the MT to the FDM potential is not as dramatic as was expected for this system, for which non-MT corrections had been long assumed as fundamental.

However the hexa-coordinate systems studied up so far do not exhibit prominent pre-edge features. A different case is the present one, a transition metal compounds with square planar coordination symmetry displaying features at the pre-edge and in the lower rising edge region, i.e. up to a photoelectron energy of about 10 eV, that are difficult to model, (Munoz-Paez et al. 2000; Chen et al. 2003). Before the present work, the MXAN method was not tested and validated for these coordination geometries, and the applicability of the method to provide structural information on these system was questioned.

However, we do not **expects to fit pre-edge features of any system accurately since it is well known that, in the calculations done in the "extended continuum" scheme, the amplitude of the absorption rate to bound states is not accurate, because the final state function is always normalized as a continuum state. Actually the main goal of the MXAN method is not to accurately fit pre-edge features, but to extract structural parameters of the compound under study by using an opportune spectral range that starts from the low-energy region of the full multiple scattering regime, i.e., the XANES continuum states, and extends up to the low-energy part of the EXAFS region**.

**Origin of peak 3**. To investigate the origin of peak **3,** we compare now in Fig. 7 the experimental and theoretical (no line-broadening) spectra by including the low-energy part excluded from the structural fit. A broad bump appears in the calculated spectrum at the position of peak **3**. The intensity of this bump depends rather strongly on the potential details and the experimental peak **3** cannot be reproduced accurately. **However we have calculated the bound-to-bound transition energies populating the pre-edge and lower rising edge region by** spin unpolarized self-consistent field (SCF-$X_\alpha$) method, imposing the formal valence of each atom (Pedio et al. 1994 and references therein), and a $D_{4h}$ symmetry of the cluster. Only the first and second shell of coordination were included in this calculation. An $X_\alpha$ MT molecular potential was built up by this approach and the absorption spectrum of the $S_0$ structure was calculated. This procedure does not allows a fitting, however it has the advantage to allow an unambiguous comparison between the calculated SCF-$X_\alpha$ transitions to empty bound states and the calculated pre-edge XANES features. According to this procedure, the broad bump in the pre-edge region belongs to the cross section polarized along the normal to the porphyrin plane which is consistent with a dipole transition from 1s to the first empty state of $A_{2u}$ symmetry. This corresponds to the 1s  $4p_z$ transition analogously to what previously assigned for square planar Cu compounds (Kau et al. 1987). **Moreover, according with our SCF-$X_\alpha$ calculations, if the nickel charge and the $D_{4h}$ symmetry are kept fixed, an increase of more than 1 eV affects the** transition energy 1s  $A_{2u}$ going from the $S_0$ to the $T_1$ state. Thus, the experimentally observed blue-shift of peak **3** should depend on the ring expansion rather than on the metal-to-ligand charge transfer involved in the S0 T1 photoexcitation. A mutual enhancement or compensation of the effects cannot be excluded. Further work would be necessary to assess if twisting effects could be also associated to energy shifts of this feature.

## Conclusions

This work demonstrates that, taking care to exclude the very low energy range, MT approximation can be used to extract structural information from K-edge XANES spectra of square-planar transition metal compounds, and difference spectra due to distortions of the coordination geometry. and confirms that XANES differential analysis can be successfully applied to understand relationships between electronic and structural states in transient photoexcited species from time-resolved experiments.

**Figure Legends**

**Fig. 1  Upper frame: Experimental Ni K-edge XANES spectrum of Ni(II)-TMP, ground state, $S_0$ (solid line), and of its 100ps laser photoproduct, containing 45-50% of an excited triplet T1 state. Lower frame: XANES difference spectrum of data reported in the upper frame**

Fig. 2. Theoretical XANES spectrum of square planar Ni(II)-porphyrin as a function of the cluster size. Calculations are done without any damping coming from inelastic losses to emphasize the origin of each feature. According to this set of calculations, peaks 4 and 5 (absent in 1-shell and 2-shell calculations) can be attributed to MS signal coming from electron pathways extending all over the 3-shell cluster.

Fig. 3 MXAN fit (solid line) of the XANES spectrum of the $S_0$ ground state (circles). We obtained $R_1$= d(Fe-Np) = (1.93±0.02) and $R_2$= d(Fe-C) = (2.94±0.03) from the fit

Fig. 4  Upper frame. XANES difference spectrum [photoproduct – $S_0$] (circles) and the MXAN best fit (solid line) giving the optimized average Ni-N and Ni-C distances, and a $T_1$ fraction of 54%.  Lower frame: XANES difference spectrum [photoproduct – $S_0$] (circles) and the MXAN best fit (solid line) giving the optimized average [Ni-pyrrol] distance and a $T_1$ fraction of  59%

Fig. 5 **Two ideal structures of the porphyrin, used for testing twisting effects on XANES. The first is perfectly planar (structure A) and the second is ruffled (structure B).**

Fig. 6. Upper frame: Theoretical XANES spectra of structure A (planar, solid line) and structure B (twisted line), having the same d(Ni-Np) distance. Lower frame: Theoretical XANES spectra of structure A with two different d(Ni-$N_p$) distances. Distance values are reported in the legend.

Fig. 7. The effect of non-coordinating solvent on the XANES spectrum of **Ni(II)-TMP**. Dots: experiment. Dotted line: Calculations done on the atomic cluster without solvent. Solid line: Calculations done with two toluene molecules added at non-coordinating distance, 4.9 Å, as described in the text. Theoretical calculations do not include line-brodening.

Table 1

**Summary of Fits of the (Photoproduct-$S_0$) XANES difference spectrum. The values found for the fit of the ground state (absolute spectrum) are: d(Ni-N)=R1=1.93±0.02; d(Ni-C)=R2=2.94±0.03**
**Fits #1-#5 are optimized in the double variable space [R1-R2], fits #6-#8 in the single variable space d(Ni-pyrrol)**
**Statistical errors are last digits in parenthesis. In fit #7 and #8 the parameter "esh" (for energy alignment exp.vs.th) is restrained to the value found for the gs fit, 8341.9±0.3**
**Notice that the T1 fraction is provided by our fit, whereas the chem.shift is assumed "a priori"; hence different fits correspond to different choices of the chemical shift**

| Fit | Chemical shift (eV) | $T_1$ fraction | $\chi^2/n$ | $\Delta$d(Ni-N) R1 (Å) | $\Delta$d(Ni-C) R2 (Å) | $\Delta$d (Ni-pyr) (Å) |
|---|---|---|---|---|---|---|
| Experiment | Chen et al. | 50% | | 0.08(2) | 0.07(2) | - |
| Fit | | | | | | |
| 1 | 0.0 | 54% | 2.21 | 0.04(2) | 0.05(3) | - |
| 2 | -0.5 | 54% | 2.36 | 0.06(2) | 0.06(3) | - |
| 3 | -1.0 | 46% | 2.86 | 0.09(2) | 0.08(5) | - |
| 4 | -1.5 | 46% | 3.57 | 0.10(2) | 0.09(4) | - |
| 5 | -2.0 | 41% | 4.06 | 0.13(3) | 0.05(4) | - |
| 6 | 0.0 | 59% | 2.56 | - | (0.05) | 0.06(2) |
| 7 ESH_fix | 0.0 | 44% | 3.45 | - | (0.06) | 0.07(2) |
| 8 ESH_fix | -0.5 | 46% | 3.10 | - | (0.07) | 0.08(2) |

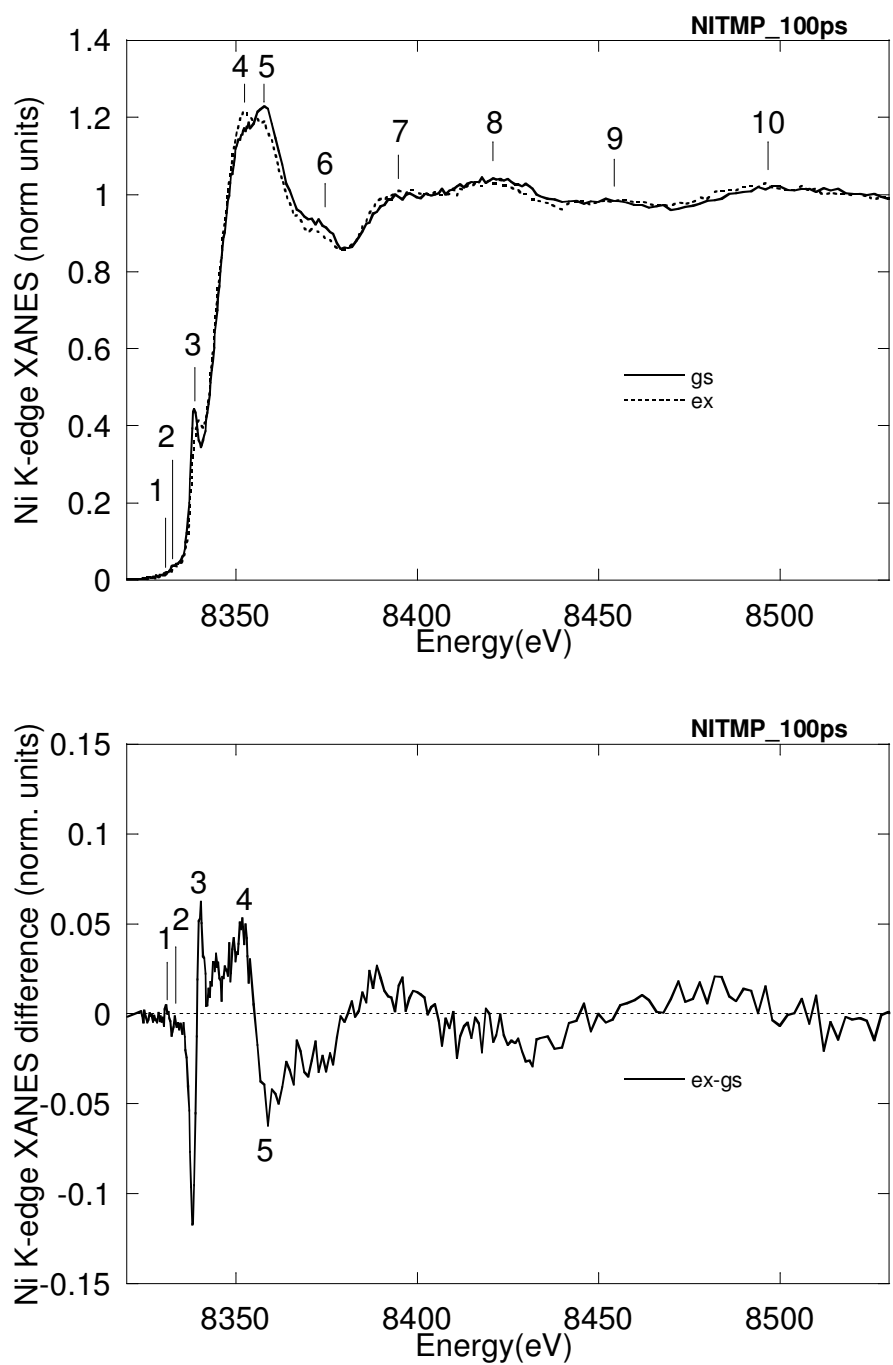

**Fig. 1** Upper frame: Experimental Ni K-edge XANES spectrum of Ni(II)-TMP, ground state, $S_0$ (solid line), and of its 100ps laser photoproduct, containing 45-50% of an excited triplet T1 state. Lower frame: XANES difference spectrum of data reported in the upper frame.

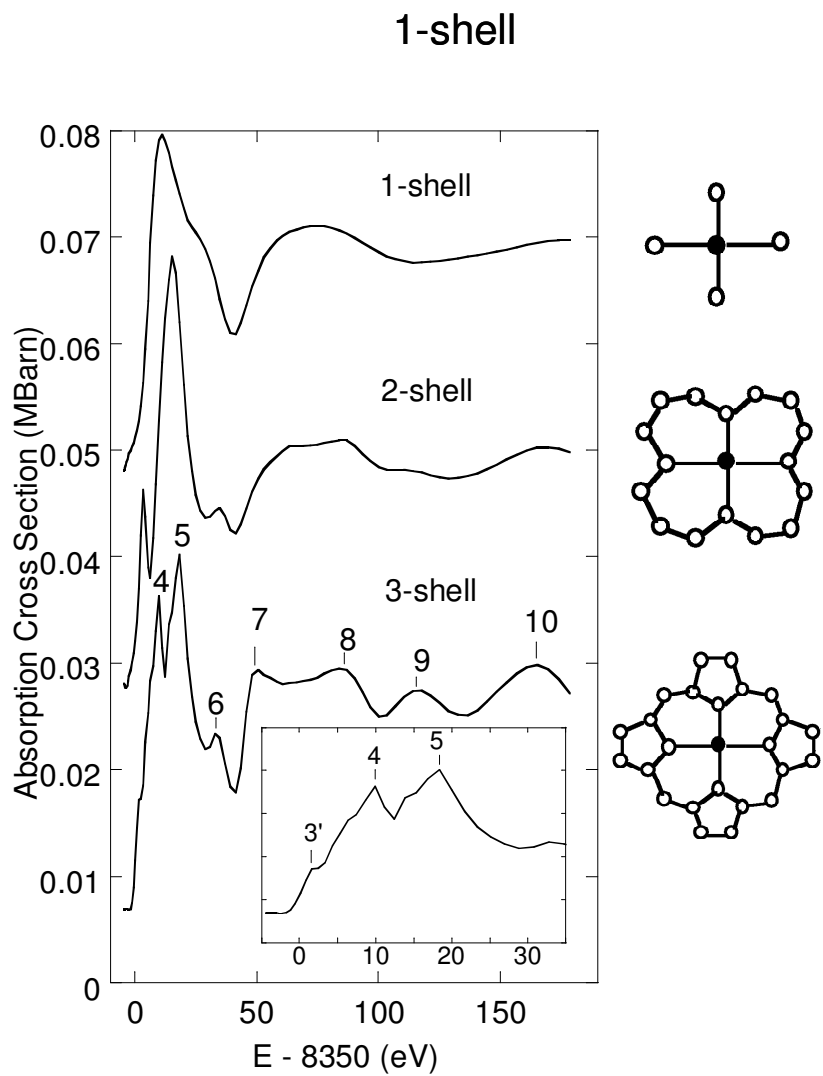

Fig. 2. Theoretical XANES spectrum of square planar Ni(II)-porphyrin as a function of the cluster size. Calculations are done without any damping coming from inelastic losses to emphasize the origin of each feature. According to this set of calculations, peaks 4 and 5 (absent in 1-shell and 2-shell calculations) can be attributed to MS signal coming from electron pathways extending all over the 3-shell cluster.

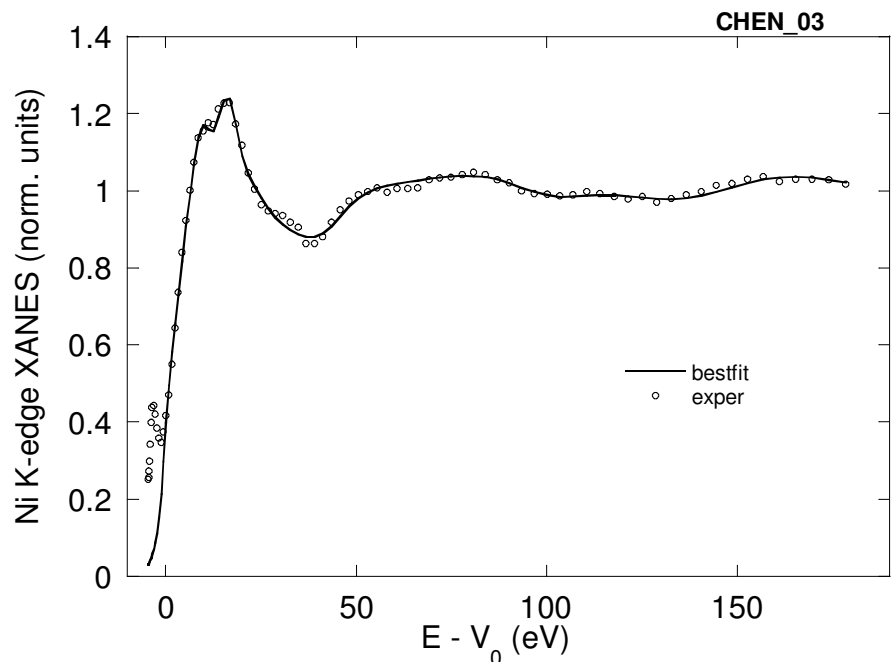

Fig. 3

MXAN fit (solid line) of the XANES spectrum of the $S_0$ ground state (circles). We obtained $R_1$= d(Fe-Np) = (1.93±0.02) and $R_2$= d(Fe-C) = (2.94±0.03) from the fit.

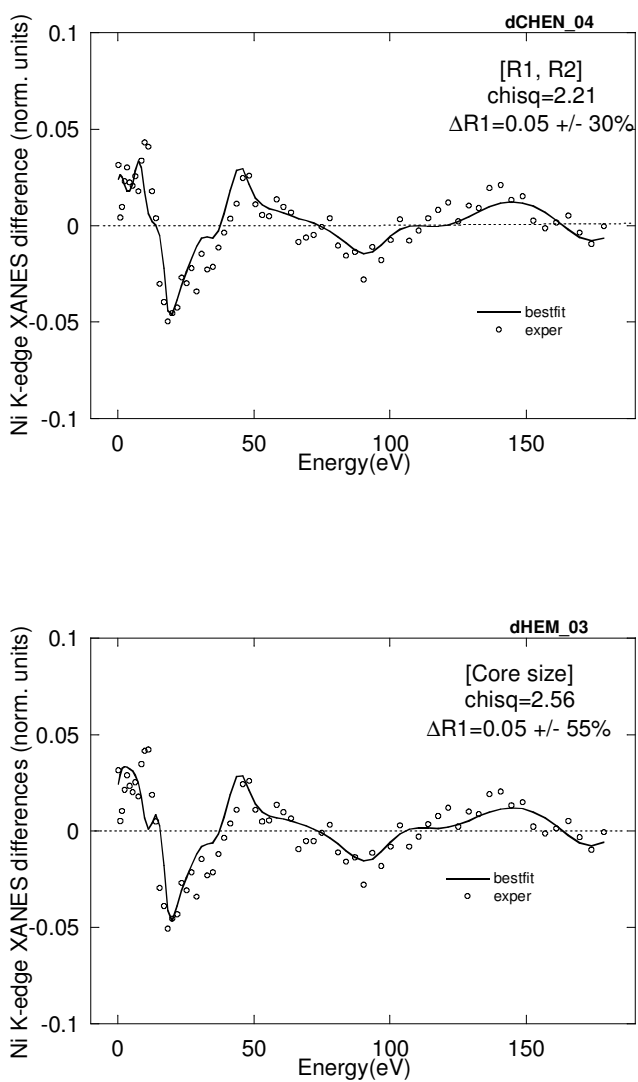

Fig. 4

Upper frame. XANES difference spectrum [photoproduct – $S_0$] (circles) and the MXAN best fit (solid line) giving the optimized average Ni-N and Ni-C distances, and a $T_1$ fraction of 54%.

Lower frame: XANES difference spectrum [photoproduct – $S_0$] (circles) and the MXAN best fit (solid line) giving the optimized average [Ni-pyrrol] distance and a $T_1$ fraction of 59%

A)

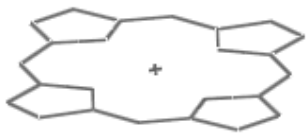

B)

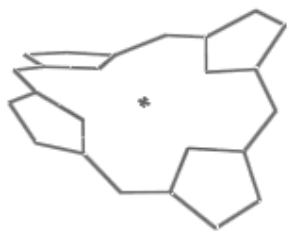

Fig. 5
Two ideal structures of the porphyrin, used for testing twisting effects on XANES. The first is perfectly planar (structure A) and the second is ruffled (structure B).

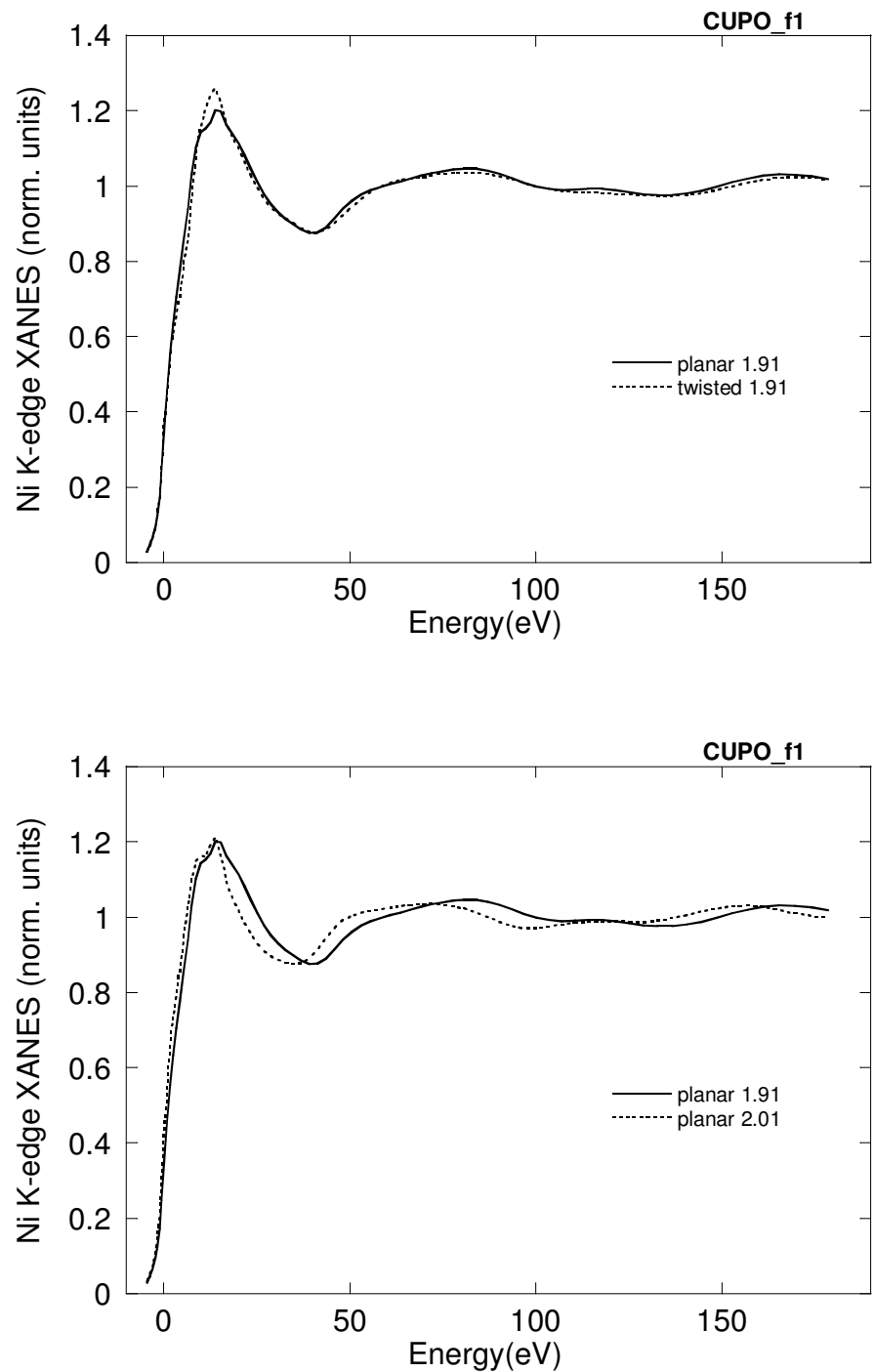

Fig. 6. Upper frame: Theoretical XANES spectra of structure A (planar, solid line) and structure B (twisted line), having the same d(Ni-Np) distance. Lower frame: Theoretical XANES spectra of structure A with two different d(Ni-$N_p$) distances. Distance values are reported in the legend.

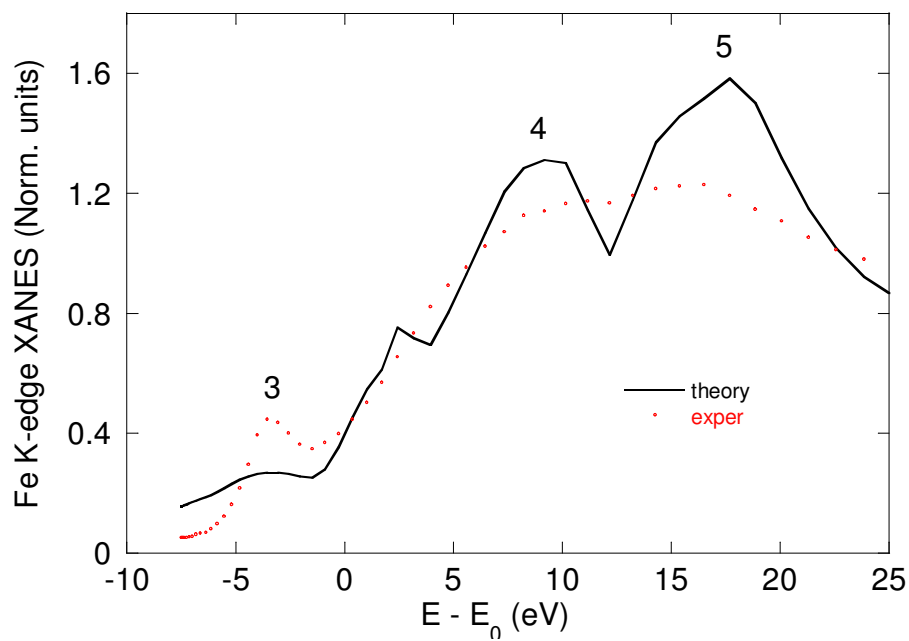

Fig. 7. The effect of non-coordinating solvent on the XANES spectrum of **Ni(II)-TMP**. Dots: experiment. Dotted line: Calculations done on the atomic cluster without solvent. Solid line: Calculations done with two toluene molecules added at non-coordinating distance, 4.9 Å, as described in the text. Theoretical calculations do not include line-brodening.